%% file: main.tex
\documentclass[twoside,final,11pt]{entics}
\usepackage{enticsmacro}

\PassOptionsToPackage{hyphens}{url}
\usepackage[hyphens]{url}
\usepackage{hyperref}
\usepackage[hyphenbreaks]{breakurl}

\usepackage[bibliography=common,appendix=append]{apxproof}

\usepackage{amsmath,amsfonts,amsthm}
\usepackage{graphicx}
\usepackage[capitalise]{cleveref}
\usepackage[inline]{enumitem}
\usepackage{ebproof}

\newtheorem*{convention}{Convention}

\newtheoremrep{theorem}{Theorem}[section]
\newtheoremrep{corollary}[theorem]{Corollary}
\newtheoremrep{proposition}[theorem]{Proposition}
\newtheoremrep{lemma}[theorem]{Lemma}
\newtheoremrep{definition}[theorem]{Definition}
\newtheoremrep{example}[theorem]{Example}

\usepackage{quiver}
\usepackage{macros}
\usepackage{hott}

\usepackage{microtype}

\def\conf{MFPS 2022} 	
\volume{1}			
\def\volu{1}			
\def\papno{1}			
\def\mydoi{{\normalshape  \href{https://doi.org/10.46298/entics.10492}{10.46298/entics.10492}}}
\def\copynames{V. Choudhury, M. Fiore} 
\def\runtitle{Free Commutative Monoids in HoTT}

\def\lastname{Choudhury and Fiore}
\begin{document}
\begin{frontmatter}
  \title{Free Commutative Monoids in Homotopy Type Theory}
  \author{Vikraman Choudhury\thanksref{email1}}
  \thanks[email1]{Email: \href{mailto:vc378@cl.cam.ac.uk} {\texttt{\normalshape vc378@cl.cam.ac.uk}}}
  \author{Marcelo Fiore\thanksref{email2}\thanksref{ack2}}
  \thanks[email2]{Email: \href{mailto:marcelo.fiore@cst.cam.ac.uk}
    {\texttt{\normalshape marcelo.fiore@cst.cam.ac.uk}}}
  \thanks[ack2]{Research partially supported by EPSRC grant EP/V002309/1.}
  \address{Department of Computer Science and Technology, University of Cambridge}

  \begin{abstract}
    \input{abstract}
  \end{abstract}

\end{frontmatter}

\input{introduction}
\input{freecommmonoids}
\input{universalconstruction}
\input{slistconstruction}
\input{generalized-logic}
\input{classicallinearlogic}
\input{differentiallinearlogic}
\input{equality}
\input{deduction-system}
\input{discussion}

\bibliographystyle{../../entics}
\bibliography{species,artifacts}

\renewcommand{\appendixsectionformat}[2]{
  {Supplementary material for Section~#1}
}


\end{document}

%% file: abstract.tex
We develop a constructive theory of finite multisets in Homotopy Type Theory,
defining them as free 
commutative monoids.
After recalling basic structural properties of the free commutative-monoid
construction, we formalise and establish the categorical universal property of
two, necessarily equivalent, algebraic presentations of free commutative
monoids using 1-HITs.  These presentations correspond to two different
equational theories invariably including commutation axioms.
In this setting, we prove important structural combinatorial properties of
finite multisets.  These properties are established in full generality without
assuming decidable equality on the carrier set.
As an application, we present a constructive formalisation of the relational
model of classical linear logic and its differential structure.  This leads to
constructively establishing that free commutative monoids are conical
refinement monoids.
Thereon we obtain a characterisation of the equality type of finite multisets
and a new presentation of the free commutative-monoid construction as a
set-quotient of the list construction.
These developments crucially rely on the commutation relation of
creation/annihilation operators associated with the free commutative-monoid
construction seen as a combinatorial Fock space.  


%% file: introduction.tex
\section{Introduction}
\label{sec:introduction}

Martin-L\"{o}f Type Theory~(MLTT) introduces the identity type to express
equality in type theory. Homotopy Type Theory and Univalent 
Foundations~(HoTT/UF)~\cite{univalentfoundationsprogramHomotopyTypeTheory2013}
embraces this proof-relevant notion of equality, and extends it with
Voevodsky's univalence principle, 
homotopy types, and Higher Inductive Types~(HITs), proposing a foundational
system for constructive mathematics.

Algebraic structures, that is, sets with operations satisfying equations, are
ubiquitous in computer science and mathematics. In HoTT/UF, given a type $A$,
and terms $x, y : A$, the identity type $x \id_{A} y$ carries interesting
structure.  Using this equality type for equations, it is a challenging
problem to define (higher) algebraic structures. In this paper, we consider
the problem of defining and studying \emph{free commutative monoids} on sets
in HoTT. 

\subsection{Finite multisets}

A multiset is intuitively a set whose elements have multiplicities. One
possible formalisation of a finite multiset on a set $A$ is simply a finite
set $S$ with elements drawn from $A$ along with a multiplicity function from
$S$ to the set of natural numbers $\Nat$.  Alternatively, and informally, this
is an \emph{unordered} list of elements drawn from $A$; that is, lists up to
reordering of elements.

How does one define finite multisets in constructive type theory? There are
several possibilities, but the basic idea is to define an equivalence relation
for permutations and quotient lists by this relation. This can be done either
using setoids or various techniques for constructing well-behaved quotients
(see~\cref{sec:Discussion} for a discussion).

Analogous to lists being free monoids, finite-multisets are free commutative
monoids.  We therefore define finite multisets in HoTT establishing the
categorical universal property of free commutative monoids.  Further, we prove
several structural properties of finite multisets without making assumptions
of decidable equality on the underlying set.

\subsection{Contributions}

The main goal of the paper is to develop a constructive theory of free
commutative monoids and their finite-multiset representation, including a
characterisation of the path space, with applications to models of
Ehrhard and Regnier's differential linear logic and combinatorial Fock space.

\begin{itemize}[leftmargin=*]
      \item
            In~\cref{sec:FreeCommutativeMonoidMonad}, we start by briefly
            defining commutative monoids and their homomorphisms in HoTT. Then,
            we state the universal property of free commutative monoids and
            describe the standard commutative monad structure of the free
            commutative monoid on a set that follows from its universal
            property. After this, we give two different constructions of the
            free commutative monoid using 1-HITs -- the standard
            universal-algebraic one and a folklore swapped-list one. We show
            that they both satisfy the universal property, making them
            equivalent.

      \item
            In~\cref{sec:generalized-logic}, we describe the power relative
            monad in HoTT.  We then develop the category of relations~$\Rel$
            in the spirit of Lawvere's Generalized Logic, where sets (hsets)
            are regarded as groupoids enriched over propositions. We exhibit
            the dagger compact structure of $\Rel$, building towards the
            relational model of Girard's linear logic.

      \item In~\cref{sec:relational-classical}, we formalise the relational
            model of classical linear logic in HoTT. We use Day's promonoidal
            convolution to lift free commutative monoids in $\Set$ to cofree
            commutative comonoids in $\Rel$ and show its Lafont linear
            exponential comonad structure.

      \item In~\cref{sec:DifferentialStructure}, we formalise the differential
            structure of the relational model of linear logic in HoTT. To do
            so, we start by establishing structural properties of subsingleton
            multisets and use this to characterise equalities between
            singleton multisets and between concatenations and projections of
            multisets. Using this we establish the differential structure
            following Fiore's axiomatisation 
            by constructing the creation operator.

      \item In~\cref{sec:MultisetEqualitySection}, we characterise the path
            space of finite multisets using the commutation relation of
            creation/annihilation operators associated with the free
            commutative-monoid construction seen as a combinatorial Fock
            space.  This leads to a sound and complete deduction system for
            multiset equality. Finally, using this relation as a path
            constructor, we introduce a new conditional-equational
            presentation for free commutative monoids.
\end{itemize}
We have a partial formalisation of our constructions and main results
(see~\cref{subsec:formalisation}). Informal paper proofs and proof sketches are
also provided in the supplementary appendices.

\subsection{Type theory}

We choose to present our results in the type-theoretic language of the HoTT
book~\cite{univalentfoundationsprogramHomotopyTypeTheory2013}. All our claims
will hold in a Cubical Type
Theory~\cite{cohenCubicalTypeTheory2018,angiuliComputationalSemanticsCartesianCubical2019,vezzosiCubicalAgdaDependently2019}
as well and our formalisation uses Cubical
Agda~\cite{vezzosiCubicalAgdaDependently2019}. We briefly remark on the
type-theoretic notation we use in the paper.

We write dependent function types as $\dfun*{x:A}{B(x)}$, or simply
${(x:A) \to B(x)}$, and dependent product types as $\dsum*{x:A}{B(x)}$, or
simply ${(x:A) \times B(x)}$, following Agda-inspired notation. Sometimes, we
leave function arguments implicit, marked with braces ${\{x:A\}}$. We also
take a few liberties with notation, when using infix or (un)curried forms of
operations, with appropriately assigned associativity.

The identity type between ${x, y : A}$ is written as $x \id_{A} y$, and we write
$p \comp q$ for path composition and $\inv{p}$ for the path inverse operations.
The constant path is given by ${\refl_{x} : x \id_{A} x}$ and
$\term{ap}_{f} : \dfun*{x,y:A}{{x \id_{A} y} \to {f(x) \id_{B} f(y)}}$ gives the
functorial action on paths.

The (univalent) universe of types is denoted $\UU$. For a type family
(fibration) ${B : A \to \UU}$, the $\term{transport}$ operation lifts paths in
the base type $A$ to functions between the fibers, that is,
$\term{transport}_{B} : \dfun*{x,y:A}{{x \id_{A} y} \to}$ ${B(x) \to B(y)}$. For
${u:B(x)}$, ${v:B(y)}$, we write $\tr{p}{u}$ for ${\transport{B}{p}{u}}$, and
${u \id^{B}_{p} v}$ for the lifted path over $p$ given by
${\tr{p}{u} \id_{B(y)} v}$. In Cubical Agda, this is given by the
heterogeneous path type $\term{PathP}$. The functorial action on sections
$f : \dfun*{x:A}{B(x)}$ of the fibration is given by
$\term{apd}_{f} : \dfun*{x,y:A}{{(p : x \id_{A} y)} \to}$ ${f(x) \id^{B}_{p} f(y)}$.

Equivalences between types $A$ and $B$ are denoted by $A \eqv B$, and homotopies
between functions $f$ and $g$ are denoted by $f \htpy g$. By
$\term{univalence}$/$\term{funext}$, these are equivalent to the corresponding
identity types. We use $\defeq$ for definitions and $\jdgeq$ for denoting
judgemental equalities.

For homotopy $n$-types, we use the standard definitions for contractible types
(-2), propositions (-1) and sets (0), which are given by ${\isContr{A}} \defeq
{\dsum*{x:A}{\dfun*{y:A}{y \id x}}}$, ${\isProp{A}} \defeq
{\dfun*{x,y:A}{\isContr{x \id y}}}$, and ${\isSet{A}} \defeq
{\dfun*{x,y:A}{\isProp{x \id y}}}$. We write $\hProp$ and $\hSet$ for the
universe of propositions and sets respectively. We write $\Set$ for the
(univalent) category of $\hSet$s and
functions~\cite[Chapter~10.1]{univalentfoundationsprogramHomotopyTypeTheory2013}.
When writing commuting diagrams, we mean that they commute up to homotopy (or
equivalently, up to the identity type).

The $n$-truncation of a type $A$ is written as $\Trunc[n]{A}$, with the point
constructor $\trunc{\blank} : A \to \Trunc[n]{A}$. When working with
propositions ($(-1)$-truncated types), we use standard logical notation, that is
truth values, $\bot \defeq \emptyt$, $\top \defeq \unit$, binary connectives
$\phi \land \psi \defeq \phi \times \psi$ and
$\phi \lor \psi \defeq \Trunc[-1]{\phi + \psi}$, and quantifiers given by
$\fora{x:A}{P(x)} \defeq \dfun*{x:A}{P(x)}$ and
$\exis{x:A}{P(x)} \defeq \Trunc[-1]{\dsum*{x:A}{P(x)}}$
(see~\cite[Definition~3.7.1]{univalentfoundationsprogramHomotopyTypeTheory2013}).


%% file: freecommmonoids.tex
\section{The free commutative-monoid monad}
\label{sec:FreeCommutativeMonoidMonad}

We start by giving the definition of commutative monoids in type theory; they
are monoids with an additional commutation axiom.

\begin{definition}[Commutative monoid]
  A commutative monoid $\monoid M = (M;\mult,e)$ is a set $M$ with a commutative monoid structure, given by:
  \begin{enumerate}
    \item
    a \emph{multiplication} function ${\blank}\mult{\blank} : M \times M \to M$;
    and
    \item a \emph{unit} element $e : M$;
  \end{enumerate}
  such that the following axioms hold:
  \begin{enumerate}
    \item for $x : M$, we have $x \mult e = x$, $e \mult x = x$;
    \item for $x, y, z : M$, we have $x \mult (y \mult z) = (x \mult y) \mult z$;
    and
    \item for  $x, y: M$, we have $x \mult y = y \mult x$.
  \end{enumerate}
\end{definition}

\begin{example}
  The natural numbers
  $\Nat$~\cite[Section~1.9]{univalentfoundationsprogramHomotopyTypeTheory2013}
  are a commutative monoid under addition, with unit~$0$, and also under
  multiplication, with unit~$1$. 
  For a set $A$, 
  lists~$\List[A]$~\cite[Section~5.1]{univalentfoundationsprogramHomotopyTypeTheory2013}
  with the empty list for unit, and the append operation for multiplication are
  a monoid, but not generally a commutative monoid.
\end{example}

\begin{definition}[Homomorphism] 
For commutative monoids $\monoid M = (M;\mult_{M},e_{M})$ and 
$\monoid N = (N;\mult_{N},e_{N})$, a function $f : M \to N$ is a
homomorphism if it preserves the unit and multiplication.
\[
\isCMonHom{f} 
\defeq 
f(e_M) \id e_N 
\ \land \ 
\forall(x, y : M). f(x \mult_M y) \id f(x) \mult_N f(y)
\enspace.
\]
The set of homomorphisms is 
$\CMonHom{\monoid M}{\monoid N} \defeq \dsum*{f:M \to N}{\isCMonHom{f}}$.
\end{definition}


\begin{definition}[Equivalence]
  Two commutative monoids $\monoid M = (M;\mult_{M},e_{M})$ and
  $\monoid N = (N;\mult_{N},e_{N})$, are equivalent as commutative monoids,
  $\monoid M \eqv_{\term{CMon}} N$, whenever there is a commutative monoid
  homomorphism $f : M \to N$ which is an equivalence:
  $\monoid M \eqv_{\term{CMon}} N \defeq \dsum*{f:M \to N}{\isCMonHom{f} \land \isEquiv{f}}$.
\end{definition}

\begin{proposition}[Structure Identity Principle for commutative monoids]
  \label{prop:sip-cmon}
  Given two commutative monoids $\monoid M = (M;\mult_{M},e_{M})$ and
  $\monoid N = (N;\mult_{N},e_{N})$, the type of their equivalences is
  equivalent to their identity type:
  $(\monoid M \eqv_{\term{CMon}} N) \eqv (M \id_{\UU} N)$.
\end{proposition}

\noindent
We are mainly interested in free commutative monoids whose definition we state
next.

\begin{definition}[Universal property of free commutative monoids]
\label{def:universal-property}
For a set $A$, a commutative monoid 
$\monoid \M[A] = (\M[A];\mappend_A,\mnil_A)$ together with a function $\eta_A:
A\to \M[A]$ is the free commutative monoid on $A$ whenever, for every
commutative monoid $\monoid M$, composition with $\eta_A$ determines an
equivalence as follows
\[
(\blank) \comp \eta_A : \CMonHom{\M[A]}{M} \xlongrightarrow{\sim} (A \to M)
\]
\noindent
Equivalently, every function $f : A \to M$ has a unique homomorphic extension
$\extend{f} : \CMonHom{\M[A]}{\monoid M}$ along $\eta_A$; that is, the type
$\dsum*{h:\CMonHom{\M[A]}{\monoid M}}{\pi_1(h) \comp \eta_A \id f}$ is 
contractible.
\end{definition}

\begin{proposition}
  \label{prop:free-cmon-nat}
  Let $\List[A]$ be the free monoid (equivalently, the list construction) on
  $A$. The free commutative monoid and the free monoid on the unit type
  $\unit$ are canonically equivalent: $\M[\unit] \eqv \List[\unit]$.  Hence,
  $\M[\unit] \eqv \Nat$.
\end{proposition}

\noindent
We have given the free commutative monoid as an endofunction $\M$ on the type
of sets $\hSet$.  It is standard that the construction extends to a strong
monad on the category of sets and functions.

\begin{definition}[Free commutative-monoid monad]
  For sets $A$ and $B$,
  \begin{enumerate}
    \item
          the unit and multiplications are $\eta_A : A\to\M[A]$ and $\mu_A
            \defeq \extend{\ps{\lam{x:A}x}} : \M[\M[A]] \to \M[A]$;

    \item
          the functorial action for a function $f: A \to B$ is $\M[f] \defeq
            \extend{\ps{\lam{a:A} \eta_B(f a)}} : \M[A] \to \M[B]$;

    \item
          the left and right tensorial strengths are
          $\sigma_{A,B}
            :  \M[A] \times B \to \M[A \times B]
            : (as,b) \mapsto \M[\lam{a:A}(a,b)](as)$,
          and
          $\tau_{A,B}
            :  A \times \M[B] \to \M[A \times B]
            : (a,bs) \mapsto \M[\lam{b:B}(a,b)](bs)$.
  \end{enumerate}
\end{definition}

\noindent
Furthermore, the free commutative-monoid monad is
commutative~\cite{kockMonadsSymmetricMonoidal1970}.
\begin{proposition} 
\label{prop:CommutativeMonad}
For all sets $A$ and $B$, 
$\extend{\tau_{A,B}}\comp\sigma_{A,\M(B)}
 \id
 \extend{\sigma_{A,B}}\comp\tau_{\M(A),B}
 : \M(A)\times\M(B)\to\M(A\times B)$.
This map is bilinear (that is, a homomorphism in each argument) and universal
amongst all such.  
\end{proposition} 

\hide{
\begin{proof}
For the identity, 
by function extensionality, we need to show that for all $as : \MSet[A]$ and
$bs : \MSet[B]$, 
${\extend{\tau}(\sigma(as,bs)) \id \extend{\sigma}(\tau(as,bs))}$. This may be
seen by induction on $as$ and $bs$ using the propositional induction
principle.
\end{proof}
}

\begin{example} \label{ex:length}
  The free commutative monoid $\M[\unit]$ on $\unit$ has the additive
  structure of natural numbers~(\cref{prop:free-cmon-nat}). Hence, any free
  commutative monoid comes equipped with a length function:
  \[
    \length_A \defeq \M[\lam{a:A}\ttt]:\M[A]\to\M[\unit]
    \enspace.
  \]
  The multiplicative structure of $\M[\unit]$ is given by
  $\eta(\ttt) : \M[\unit]$ and
  $\extend{ \pi_2 } \comp \sigma_{\unit,\M[\unit]}
    : \M[\unit] \times \M[\unit] \to \M[\unit]$.
  This is commutative by \cref{prop:CommutativeMonad}.
\end{example}

\begin{propositionrep}
  \label{prop:SeelyIso}
  The free commutative-monoid monad is canonically a strong symmetric monoidal
  endofunctor from the coproduct monoidal structure to the product monoidal
  structure; the monoidal isomorphisms are
  \[\begin{tikzcd}[ampersand replacement = \&,
        column sep = 2.5em, row sep = 0em]
      {\M(A) \times \M(B)} \&\&\& {\M(A+B)}
      \\
      \&\&
      {\M(A+B) \times \M(A+B)}
      \&
      \arrow["\simeq", from=1-1, to=1-4]
      \arrow["{\M(\imath_1)\times\M(\imath_2)}"', from=1-1, to=2-3]
      \arrow["{\mappend_{(A+B)}}"', from=2-3, to=1-4]
    \end{tikzcd}
    \ ,\enspace
    \begin{tikzcd}[ampersand replacement = \&]
      {\unit} \&\&  {\M(\nulltype)}
      \arrow["{\lam{x:\unit}\!\!\mnil_\nulltype}"', from=1-1, to=1-3 , "\simeq"]
    \end{tikzcd}\]
\end{propositionrep}

\begin{proof}
  Being the product of two commutative monoids, $\M[A] \times \M[B]$ is a
  commutative monoid and one obtains the inverse 
  $\M[A + B] \to \M[A] \times \M[B]$ of the given map 
  $\M[A] \times \M[B] \to \M[A + B]$ by homomorphically extending the map
  \[
  \big[ 
     \ \lam{a:A}(\,\eta(a),\mnil\,) 
     \ \mid 
     \ \lam{b:B}(\,\mnil,\eta(b)\,) \ 
  \big]
  : A + B \to \M[A] \times \M[B] 
   \enspace.
  \]
  On the other hand, the unique map $\M[\nulltype]\to\unit$ is the inverse of
  the given map $\unit\to\M[\nulltype]$.
\end{proof}

We have described the universal property of free commutative monoids but we
have not yet given a construction for them. We will describe
three constructions in HoTT:
\begin{enumerate*}
  \item the naive one from universal
  algebra~(\cref{subsec:UniversalAlgebraConstruction});
  \item a folklore swapped-list construction from computer
  science~(\cref{subsec:FiniteMultisetConstruction}); and
  \item a new quotiented-list construction (\cref{subsec:DeductionSystem})
  stemming from our study of multiset equality
  (\cref{subsec:CommutationRelation}).
\end{enumerate*}

The situation concerning the constructions (i) and (ii) is analogous to that
of free monoids, which can either be described by generators and relations or
by using the inductive list type
(see~\cite[Chapter~6.11]{univalentfoundationsprogramHomotopyTypeTheory2013});
our construction (iii) is of independent interest.
For each construction, our goal is to establish the categorical universal
property; so that they automatically acquire the structure detailed in this
section, albeit here established in an implementation-independent manner.
This is a distinguishing feature of our work compared to previous approaches
to finite multisets in type theory.


%% file: universalconstruction.tex
\subsection{Universal-algebraic construction}
\label{subsec:UniversalAlgebraConstruction}

Using the standard construction of free algebras for equational theories from
universal algebra, the free commutative monoid on a set can be defined as
follows.

\begin{definition}[$\FCM$]
  \label{def:fcm}
  Let $A$ be a type, we define the 1-HIT $\FCM[A]$ with the following point and
  path constructors:

  \begin{align*}
    \eta                  & : A \to \FCM[A]                                                       \\
    e                     & : \FCM[A]                                                             \\
    {\blank}\mult{\blank} & : \FCM[A] \times \FCM[A] \to \FCM[A]                                  \\
    \assoc                & : (x\;y\;z : \FCM[A]) \to x \mult (y \mult z) \id (x \mult y) \mult z \\
    \unitl                & : (x : \FCM[A]) \to e \mult x \id x                                   \\
    \unitr                & : (x : \FCM[A]) \to x \mult e \id x
                                                                                                  \\
    \comm                 & : (x\;y : \FCM[A]) \to x \mult y \id y \mult x                        \\
    \truncc               & : \isSet{\FCM[A]}
  \end{align*}
\end{definition}

\noindent
The first constructor $\eta$ postulates that $A$ maps to $\FCM[A]$. The next two give the operations on a monoid: the
identity element and multiplication (written in an infix style). The four path constructors after that assert the axioms
of a commutative monoid -- associativity, unitality, and commutativity. Finally, we add a path constructor to truncate
$\FCM[A]$ to a set, since we wish to ignore any higher paths introduced by the path constructors. For completeness, we
state the induction principle for $\FCM$ (\cref{def:fcm-induction}).

\begin{toappendix}
\begin{definition}[Induction principle for $\FCM$]
  \label{def:fcm-induction}
  Given a type family ${P : \FCM[A] \to \UU}$ with the following data:

  \begin{align*}
    \eta^{*}                  & : (x : A) \to P(\eta(x))                                                            \\
    e^{*}                     & : P(e)                                                                              \\
    {\blank}\mult^{*}{\blank} & : \{x\;y : \FCM[A]\} \to P(x) \to P(y) \to P(x \mult y)                               \\
    \assoc*                   & : \{x\;y\;z : \FCM[A]\} (p : P(x)) (q : P(y)) (r : P(z))                             
                              \to p \mult^{*} (q \mult^{*} r) \id_{\assoc[x,y,z]}^{P} (p \mult^{*} q) \mult^{*} r \\
    \unitl*                   & : \{x : \FCM[A]\} (p : P(x)) \to e^{*} \mult^{*} p \id_{\unitl[x]}^{P} p              \\
    \unitr*                   & : \{x : \FCM[A]\} (p : P(x)) \to p \mult^{*} e^{*} \id_{\unitr[x]}^{P} p              \\
    \comm*                    & : \{x\;y : \FCM[A]\} (p : P(x)) (q : P(y))                                       
                              \to p \mult^{*} q \id_{\comm[x,y]}^{P} q \mult^{*} p                                \\
    \truncc*                  & : (x : \FCM[A]) \to \isSet{P(x)}
  \end{align*}

  there is a unique function $f : (x : \FCM[A]) \to P(x)$ with the following computation rules:

  \begin{align*}
    f(\eta(x))             & \jdgeq \eta^{*}(x)                \\
    f(e)                   & \jdgeq e^{*}                      \\
    f(x \mult y)           & \jdgeq f(x) \mult^{*} f(y)        \\
    \apd{f}{\assoc[x,y,z]} & \id \assoc*[f(x),f(y),f(z)] \\
    \apd{f}{\unitl[x]}     & \id \unitl*[f(x)]               \\
    \apd{f}{\unitr[x]}     & \id \unitr*[f(x)]               \\
    \apd{f}{\comm[x,y]}    & \id \comm*[f(x),f(y)]
  \end{align*}

  Note that we use judgemental equality for the computation rules for the
  point constructors and the computation rules for the path constructors hold
  up to the identity type.
\end{definition}
\end{toappendix}

$\FCM[A]$ is straightforwardly a commutative monoid. As expected, using the
induction principle and computation rules, one proves that it satisfies the
universal property of the free commutative
monoid~(\cref{def:universal-property}).

\begin{theoremrep}
  For every set $A$, $\eta:A\to\FCM[A]$ is the free commutative monoid on $A$.
\end{theoremrep}

\begin{proof}
  Given $f : A \to M$, we define $\extend{f} : \FCM[A] \to M$ by induction. It is straightforward to check that
  $\extend{f}$ is a 
  homomorphism. Finally, we need show that $\extend{(-)}$ is inverse to $(-) \comp \eta$. We establish the homotopies
  $\extend{(h \comp \eta)} \htpy h$ and $\extend{f} \comp \eta \htpy f$ by calculation.
\end{proof}


%% file: slistconstruction.tex
\subsection{Swapped-list construction}
\label{subsec:FiniteMultisetConstruction}

Alternatively, one can define free commutative monoids by means of a
\emph{folklore} swapped-list construction~(also presented
in~\cite{choudhuryFinitemultisetConstructionHoTT2019}).  Analogous to free
monoids being given by finite lists (defined as an inductive type), free
commutative monoids can be defined as lists, but identified up to swappings.

\begin{definition}[$\MSet$]
  \label{def:mset}
  The \emph{swapped-list} over a type $A$ is given by the type $\MSet[A]$,
  which is the 1-HIT with the following point and path constructors:
  \begin{align*}
    \nil                  & : \MSet[A]                       \\
    {\blank}\cons{\blank} & : A \times \MSet[A] \to \MSet[A] \\
    \swap                 & : (x\;y : A) (xs : \MSet[A])
    \to x \cons y \cons xs \id y \cons x \cons xs            \\
    \truncc               & : \isSet{\MSet[A]}
  \end{align*}
\end{definition}

The first two constructors are the standard ones for lists (free monoids), that
is, $\nil$ and $\term{cons}$ or {(infix)~$\cons$}, but we have an additional
$\swap$ path constructor asserting that the first two elements of any list can
be swapped around. Note that, this is a recursively defined HIT, where the
points and paths are recursively generated. In particular, by applying
$\term{cons}$ on paths by congruence and path composition, these adjacent swaps
can be performed anywhere in the list. Finally, we truncate $\MSet[A]$ to a
set, since we wish to ignore the higher paths introduced by $\swap$.

Classically, finite multisets are usually defined as lists quotiented by
permutations of their elements. In our setting, we generate this equivalence
relation recursively, that is, we are generating the path space of finite
multisets simply using the $\swap$ path constructor. To establish this fact
formally, we prove that $\MSet[A]$ is indeed the free commutative monoid on
$A$ by establishing its universal property.

The main means of reasoning about swapped-lists is its induction principle
(\cref{def:mset:ind}). It is akin to the induction principle for lists, but
one needs to additionally enforce invariance under swapping. Also, since we
truncate to sets, one is only allowed to eliminate to sets. This is in general
restrictive. However, for the purposes of this paper, we are only interested
in establishing the universal property of free commutative monoids and only
ever need to eliminate to sets (or propositions).

\begin{toappendix}
  \begin{definition}[Induction principle for $\MSet$]
    \label{def:mset:ind}
    \leavevmode\\
    Given a type family ${P : \MSet[A] \to \UU}$ with the following data:
    \begin{align*}
      \nil*                  & : P(\nil)                                                     \\
      {\blank}\cons*{\blank} & : (x : A) \{xs : \MSet[A]\} \to P(xs) \to P(x \cons xs)       \\
      \swap*                 & : (x\;y : A) \{xs : \MSet[A]\} (p : P(xs))
      \to\;                  x \cons* y \cons* p \id_{\swap[x,y,xs]}^{P} y \cons* x \cons* p \\
      \truncc*               & : (xs : \MSet[A]) \to \isSet{P(xs)}
    \end{align*}
    there is a unique function $f : {(xs : \MSet[A])} \to P(xs)$ satisfying the following computation rules:
    \begin{align*}
      f(\nil)                & \jdgeq nil*           \\
      f(x \cons xs)          & \jdgeq x \cons* f(xs) \\
      \apd{f}{\swap[x,y,xs]} & \id \swap*[x,y,f(xs)]
    \end{align*}
  \end{definition}
\end{toappendix}

When eliminating to propositions, the induction principle can be simplified
further. One only needs to provide an image for the $\nil$ and $\cons$
constructors, similar to the case for lists.

\begin{lemmarep}[Propositional induction principle for $\MSet$]
  \label{lem:mset:prop-ind}
  \leavevmode\\
  Let $P : \MSet[A] \to \UU$ be a type family 
  with the following data.
  \begin{align*}
    \nil*                  & : P(\nil)                                               \\
    {\blank}\cons*{\blank} & : (x : A) \{xs : \MSet[A]\} \to P(xs) \to P(x \cons xs) \\
    \truncc*               & : (xs : \MSet[A]) \to \isProp{P(xs)}
  \end{align*}
  Then, there is a unique function 
  $
{(xs : \MSet[A])} \to P(xs)$ 
  satisfying appropriate computation rules.
\end{lemmarep}

\begin{proof}
  Using the standard induction principle for $\MSet$, we need give an image
  $\swap*$ for $\swap$. Consider $\phi_{1} : P(x \cons y \cons xs)$ and
  $\phi_{2} : P(y \cons x \cons xs)$. Transporting $\phi_{1}$ along $\swap$, and
  using the fact that they are propositions, one gets the path $\phi_{1}
    \id_{\swap[x,y,xs]}^{P} \phi_{2}$.
\end{proof}

\noindent
We now exhibit the swapped-list construction as the free commutative-monoid
construction.

\begin{definition}[$\eta$]
  The generators map is given by
  $\eta : A \to \MSet[A] : a \longmapsto \sing{a} \defeq (a \cons \nil)$.
\end{definition}

\noindent
We show that $\MSet[A]$ is a commutative monoid with the empty multiset $\nil$
as unit and the binary multiplication given by the concatenation (or append)
operation $\append$ defined below.

\begin{definition}[$\append$]
  The \emph{concatenation operation}
  ${\blank}\append{\blank} : \MSet[A] \to \MSet[A] \to \MSet[A]$
  is defined by recursion on the first argument. Using the recursion principle
  on the point and path constructors, we have
  \begin{align*}
    \nil \append ys                      & \defeq ys                       \\
    (x \cons xs) \append ys              & \defeq x \cons (xs \append ys)  \\
    \ap{(\blank)\append ys}{\swap[x,y,xs]} & \defeq \swap[x,y,xs \append ys]
  \end{align*}
  Finally, $\MSet[A]$ is a set and so is $\MSet[A] \to \MSet[A]$, hence it respects $\truncc$.
\end{definition}

\noindent
To show that $\MSet[A]$ is a commutative monoid, one needs to prove a few
identities about elements of $\MSet[A]$. Since $\MSet[A]$ is a set, its
equality type is a proposition, so one can use the propositional induction
principle from~\cref{lem:mset:prop-ind}. Hence, the proofs of the monoid laws
are simply the ones for lists.

\begin{lemmarep}
  The concatenation operation $\append$ is associative and $\nil$ is a left and
  right unit.  For all $xs, ys, zs : \MSet[A]$, we have ${xs \append (ys \append
        zs)} \id {(xs \append ys) \append zs}$, $\nil \append xs \id xs$, and $xs
    \append \nil \id xs$.
\end{lemmarep}

\begin{proof}
  By the propositional induction principle, the only required cases are for
  $\nil$ and $\cons$, and this is the same proof as for lists.
\end{proof}

\noindent
We further need to establish that the concatenation operation is commutative.
Similar to proving commutativity of addition for natural numbers, this is
shown in steps as follows.
For details, we refer the reader to the supplementary material
in~\cref{lem:app-comm-apx}.

\begin{lemmarep}
  \label{lem:app-comm}
  For ${x : A}$ and ${xs, ys : \MSet[A]}$, we have
  ${x \cons xs} \id {xs \append \sing{x}}$, and ${xs \append ys} \id {ys \append xs}$.
\end{lemmarep}

\begin{proof}
  The idea is to recursively apply the $\swap$ constructor to shift the first
  element to the end. To prove commutativity by induction, the multiset is first
  split into $\term{append}$s of singleton multisets.
  \begin{enumerate}
    \item
          By induction on $xs$, we have
          \begin{align*}
            x \cons \nil
             & \id \nil \append \sing{x} & \text{left unit}
          \end{align*}
          and
          \begin{eqproof}
            & x \cons (y \cons xs) & \\
            \id & y \cons (x \cons xs) & \text{by } \swap[x,y,xs] \\
            \id & y \cons (xs \append \sing{x}) & \text{induction hypothesis} \\
            \jdgeq & (y \cons xs) \append \sing{x} & \text{definition}
          \end{eqproof}

    \item
          By induction on $xs$, we have
          \begin{eqproof}
            & \nil \append ys & \\
            \jdgeq & ys & \text{definition} \\
            \id    & ys \append \nil & \text{right unit}
          \end{eqproof}
          and
          \begin{eqproof}
            & (x \cons xs) \append ys & \\
            \jdgeq & x \cons (xs \append ys) & \text{definition} \\
            \id    & x \cons (ys \append xs) & \text{induction hypothesis} \\
            \jdgeq & (x \cons ys) \append xs & \text{definition} \\
            \id    & (ys \append \sing{x}) \append xs & \text{by (i)} \\
            \id    & ys \append (\sing{x} \append xs) & \text{associativity} \\
            \jdgeq & ys \append (x \cons xs)  & \text{definition} \\
          \end{eqproof}
  \end{enumerate}
\end{proof}


\begin{toappendix}
\begin{proposition}
  \label{prop:slist-extend}
  Given a set $A$, a commutative monoid $M$, and a map $f : {A \to M}$ of sets, $f$ extends to a homomorphism
  ${\extend{f} : \CMonHom{\MSet[A]}{M}}$.
\end{proposition}

\begin{proof}
  We define $\extend{f} : \MSet[A] \to M$ by induction. On the point
  constructors, we let
  \begin{align*}
    \extend{f}(\nil)       & \defeq e                         \\
    \extend{f}(x \cons xs) & \defeq f(x) \mult \extend{f}(xs)
  \end{align*}
  On the path constructor, we have
  \begin{eqproof}
    & f(x) \mult (f(y) \mult \extend{f}(xs)) & \\
    \id & (f(x) \mult f(y)) \mult \extend{f}(xs) & \text{associativity} \\
    \id & (f(y) \mult f(x)) \mult \extend{f}(xs) & \text{commutativity} \\
    \id & f(y) \mult (f(x) \mult \extend{f}(xs)) & \text{associativity}
  \end{eqproof}
  To check that this is a 
  homomorphism, we establish that
  $\extend{f}(xs \append ys) \id \extend{f}(xs) \mult \extend{f}(ys)$ for all $xs, ys : \MSet[A]$. Since this is a
  proposition, by induction on $xs$, we have
  \begin{eqproof}
    & \extend{f}(\nil \append ys) & \\
    \jdgeq & \extend{f}(ys) & \text{definition} \\
    \id & e \mult \extend{f}(ys) & \text{unit law} \\
    \jdgeq & \extend{f}(\nil) \mult \extend{f}(ys) & \text{definition} \\
  \end{eqproof}
  \begin{eqproof}
    & \extend{f}((x \cons xs) \append ys) & \\
    \jdgeq & \extend{f}(x \cons (xs \append ys)) & \text{definition} \\
    \jdgeq & f(x) \mult \extend{f}(xs \append ys) & \text{definition}  \\
    \id & f(x) \mult (\extend{f}(xs) \mult \extend{f}(ys)) & \text{induction hypothesis} \\
    \id & (f(x) \mult \extend{f}(xs)) \mult \extend{f}(ys) & \text{associativity} \\
    \jdgeq & \extend{f}(x \cons xs) \mult \extend{f}(ys) & \text{definition} \\
  \end{eqproof}
\end{proof}
\end{toappendix}

\noindent
Finally, one proves that $\MSet$ satisfies the universal property of free
commutative monoids (\cref{def:universal-property}).

\begin{theoremrep}
  For every set $A$, $\eta:A\to\MSet[A]$ is the free commutative monoid on $A$.
\end{theoremrep}

\begin{proof}
  We get the $\extend{(-)}$ operation by~\cref{prop:slist-extend}. We need to
  show that this is inverse to $(-) \comp \eta$. We establish the homotopies
  $\extend{(h \comp \eta)} \htpy h$ and $\extend{f} \comp \eta \htpy f$ by
  calculation.
\end{proof}

\noindent
As both $\FCM$ and $\MSet$ satisfy the same categorical universal property, it
follows that they are equivalent as commutative monoids.

\begin{corollaryrep}
  For every set $A$, we have an equivalence \( \MSet[A] \eqv_{\type{CMon}} \FCM[A] \).
\end{corollaryrep}

\begin{proof}
  The two maps are constructed by extending $\eta$; that is, as
  $\extend{\eta}$.  Using the universal property, their composition is
  homotopic to $\idfunc$.
\end{proof}

\begin{convention}
  We will henceforth use $\M[A]$ to indicate the 
  \emph{free commutative monoid} on a set $A$ that will be also referred to as
  the \emph{finite-multiset construction}.
\end{convention}

\noindent
Our results for the free commutative-monoid construction $\M$ will therefore
apply to either $\FCM$ or $\MSet$ using $\term{SIP}$ (\cref{prop:sip-cmon}) and
$\term{transport}$.


%% file: generalized-logic.tex
\section{Generalized logic}
\label{sec:generalized-logic}

This section, which stands in its own right, is an interlude to consider the
category of relations in
HoTT~\cite[Example~9.1.19]{univalentfoundationsprogramHomotopyTypeTheory2013}.
This is pivotal in our subsequent study and applications of the free
commutative-monoid construction to be carried out in the rest of the paper.

We consider HoTT from the viewpoint of Lawvere's Generalized
Logic~\cite{lawvereMetricSpacesGeneralized1973a}. The latter takes place in
enriched category theory and the analogy we pursue here is that of regarding
$\hSet$s as groupoids \emph{intrinsically} (weakly) enriched over propositions
(or, in informal technical jargon, $\hProp$-groupoids with the groupoid
structure provided by the identity type, with $\hProp$ considered with its
complete Heyting algebra structure). Central to us, is that this turns out to
provide the appropriate framework for developing the calculus of relations in
HoTT.

\subsection{Power objects}
\label{subsec:PowerObjects}

We begin by introducing power objects.  Recalling that we work in predicative
type theory, we use subscripts to indicate universe levels.

\begin{definition}[Power construction] We define
  $\PSet : \hSet_i \to \hSet_{i+1} : A \longmapsto (A \to \hProp_i)$.
\end{definition}

\noindent
Note that the above is not a \emph{powerset} construction -- it is not even an
endofunction on $\hSet_{i}$. If so inclined, one could turn it into the powerset
monad using Voevodsky's \emph{propositional resizing
  axiom}~\cite{voevodskyResizingRulesTheir2011} to lower the universe level.
Instead, we choose here to work with the power construction as a \emph{relative
  monad}~\cite{altenkirchMonadsNeedNot2010,fioreRelativePseudomonadsKleisli2018}
and follow the framework of Kleisli
bicategories~\cite{hylandElementsTheoryAlgebraic2014,fioreRelativePseudomonadsKleisli2018}
(restricted 1-categorically) to consider relations in HoTT. Henceforth, we
drop the universe levels.

From the viewpoint of generalized logic, the power relative monad structure is
given by the Yoneda embedding and left Kan extension along it. Internally, in
HoTT, these operations are easily constructed using the identity type and
(suitably truncated) dependent product types, respectively.

\begin{definition}[Power relative monad]
  We define the relative monad operations for $\PSet$ as follows.
  \begin{enumerate}
    \item
          The unit is 
          $\yo_A : A \to \PSet[A] : a \longmapsto \lam{x:A} a \id_{A} x$.

    \item
          The extension operation, for $f : A \to \PSet[B]$, is $\lift{f} :
            \PSet[A] \to \PSet[B] : (\alpha,b) \longmapsto \exis{a:A}{f(a,b) \land
              \alpha(a)}$.
  \end{enumerate}
\end{definition}

\noindent
We establish the following properties of the Kleisli structure which are used
later in~\cref{prop:RelCategory}.

\begin{propositionrep}
  \label{prop:kleisli-structure}
  The following identities hold.
  \begin{enumerate}
    \item \( \lift{\yo_A} \id \idfun[\PSet[A]] \).
    \item For \( f : A \to \PSet[B] \), we have \( f \id \lift{f} \comp \yo_A \).
    \item For \( f : A \to \PSet[B] \) and \( g : B \to \PSet[C] \), we have
          \( \lift{(\lift{g} \comp f)} \id \lift{g} \comp \lift{f} \).
  \end{enumerate}
\end{propositionrep}

\begin{proof}
  \leavevmode
  \begin{enumerate}
    \item For any $f : A \to \PSet[A]$ and $a : A$, we have
          \begin{align*}
            \lift{\yo_A}(f,a)
             & \jdgeq \exis{b:A}{\yo_A(a,b) \times f(a)}   \\
             & \jdgeq \exis{b:A}{\ps{a \id b} \times f(a)} \\
             & \eqv f(a)
          \end{align*}
    \item For any $a : A$ and $b : B$, we have
          \begin{align*}
            \lift{f}(\yo_A(a),b)
             & \jdgeq \exis{c:A}{f(c,b) \times \yo_A(a,c)}   \\
             & \jdgeq \exis{c:A}{f(c,b) \times \ps{a \id c}} \\
             & \id f(a,b)
          \end{align*}
    \item For any $\alpha : \PSet[A]$ and $c : C$, we have
          \begin{eqproof}
            & \lift{(\lift{g} \comp f)}(\alpha,c) & \\
            \jdgeq & \exis{x:A}{(\lift{g} \comp f)(x,c) \times \alpha(x)} & \\
            \jdgeq & \exis{x:A}{\lift{g}(f(x))(c) \times \alpha(x)} & \\
            \jdgeq & \exis{x:A}{\exis{y:B}{\ps{g(y,c) \times f(x,y)} \times \alpha(x)}} & \\
            \eqv   & \exis{y:B}{g(y,c) \times (\exis{x:A}{f(x,y) \times \alpha(x)})} & \\
            \jdgeq & \exis{y:B}{\ps{g(y,c) \times \lift{f}(\alpha)(y)}} & \\
            \jdgeq & \lift{g}(\lift{f}(\alpha))(c) & \\
            \jdgeq & (\lift{g} \comp \lift{f})(\alpha,c) & \\
          \end{eqproof}
  \end{enumerate}
\end{proof}

\subsection{The category of relations}
\label{subsec:CategoryRel}

We define the category of
relations~\cite[Example~9.1.19]{univalentfoundationsprogramHomotopyTypeTheory2013})
as the Kleisli category of $\PSet$.

\begin{definition}[$\Rel$]
$\Rel$ has objects $\hSet$s and homs \( A \proarrow B \defeq A \to \PSet[B] \).
\end{definition}


\begin{propositionrep}
  \label{prop:RelCategory}
  $\Rel$ is a (univalent) category.
\end{propositionrep}

\begin{proof}
  We note that $\Hom_{\Rel}(A,B) \defeq A \proarrow B$ is an $\hSet$ since
  $\hProp$ is. The identity arrow in $\Hom(A,A)$ is given by $\yo_A$. The
  composition of $g : B \proarrow C$ and $f : A \proarrow B$ is given by
  $\lift{g} \comp f$. The unit and associativity laws follow
  from~\cref{prop:kleisli-structure}. To show that $\Rel$ is univalent,
  see~\cite[Example~9.1.19]{univalentfoundationsprogramHomotopyTypeTheory2013}.
\end{proof}

\noindent
The category $\Rel$ has a symmetric \emph{tensor} monoidal structure given by
the cartesian product of sets $A \otimes B \defeq A \times B$ with unit
$\unit$.
It is also closed, with ${A \multimap B} \defeq A \times B$. The tensor-hom
adjunction is given by:
\[
  \begin{array}{rclll}
    C\!\otimes\! A \!\proarrow\! B
     & \jdgeq (C \!\times\! A) \!\to\! B \!\to\! \hProp
     & \simeq C \!\to\! A \!\to\! B \!\to\! \hProp
     & \simeq C \!\to\! (A \!\times\! B) \!\to\! \hProp
     & \jdgeq C \!\proarrow\! (A \!\multimap\! B)
    \enspace.
  \end{array}
\]
\noindent
In fact, $\Rel$ is compact
closed~\cite{dayNoteCompactClosed1977,kellyCoherenceCompactClosed1980}, as the
canonical map \(  \kappa : {C\otimes (A\multimap B)} \longrightarrow
{A\multimap(C\otimes B)} \) is an equivalence.
Furthermore, as $A^{\bot} \jdgeq A\multimap \unit \eqv A$, $\Rel$ is self
dual, with involution given by
\[
  (-)^\dagger : A\proarrow B
  \jdgeq A\to(B\to\hProp)
  \simeq B\to(A\to\hProp)
  \jdgeq B\proarrow A
  \enspace.
\]
To summarise:
\begin{proposition} \label{prop:RelDaggerCompact}
  The category $\Rel$ is dagger compact.
\end{proposition}

\begin{definition}
  We define the inclusion $(-)_\ast : \Set \to \Rel$ as mapping functions
  $f : A \to B$ to relations $\yo_{B} \comp f : A \proarrow B$.
\end{definition}

Note that $(-)_\ast$ preserves coproducts, which in $\Rel$ become biproducts.
Therefore, as is well known, every set is endowed with a biproduct commutative
bialgebra structure; namely, compatible coproduct commutative monoid and
product commutative comonoid structures, crucially satisfying the following
identity:
\begin{equation}\label{BiproductBialgebraLaw}
  \begin{tikzcd}[column sep = 3em, row sep = 1.5em]
    A+A & A & A+A
    \\
    A+A+A+A & & A+A+A+A
    \arrow["{\nabla}", "\shortmid"marking, from=1-1, to=1-2]
    \arrow["{\Delta}", "\shortmid"marking, from=1-2, to=1-3]
    \arrow["{\Delta+\Delta}"swap, "\shortmid"marking, from=1-1, to=2-1]
    \arrow["{\nabla+\nabla}"swap, "\shortmid"marking, from=2-3, to=1-3]
    \arrow["{\idfun[\M(A)]+c+\idfun[\M(A)]}"swap, "\shortmid" marking,
    from=2-1, to=2-3]
  \end{tikzcd}
\end{equation}
where $c\defeq[\imath_2,\imath_1]$ is the symmetry isomorphism.





%% file: classicallinearlogic.tex
\section{Relational model of linear logic}
\label{sec:relational-classical}

We give a formalisation of the \emph{relational model of linear logic} (see,
in particular, Hyland~\cite[Section~3]{hylandReasonsGeneralisingDomain2010})
developing it in HoTT from the perspective of generalized logic considered in
the previous section.
We establish that the compact closed category with finite biproducts $\Rel$
admits cofree commutative comonoids and is therefore a
model of Girard's linear logic with linear exponential comonad
structure~\cite{bentonLinearLcalculusCategorical1993,hylandGlueingOrthogonalityModels2003a}
as first considered by Lafont~\cite{lafontLinearAbstractMachine1988}.

The key to the development is the lifting of (commutative) monoid structure to
power objects. This is achieved by means of Day's \emph{promonoidal
  convolution}~\cite{dayClosedCategoriesFunctors1970}.

\hide{
  \begin{definition}[Convolution]
    If $(M,e,\mult)$ is a (commutative) monoid, then $\PSet(M)$ is a
    (commutative) monoid with unit $\hat{e}$ and multiplication $\hat{\mult}$
    given by
  \end{definition}
}

\begin{theorem}[Promonoidal convolution]
  \label{thm:promonoidal-convolution}
  If \( m : M\otimes M \proarrow M \leftproarrow \unit : e \) is a
  (commutative) monoid in $\Rel$, then $\PSet(M)$ is a (commutative) monoid in
  $\Set$, with multiplication $\hat{m}:{\PSet(M)\times\PSet(M)}\to\PSet(M)$
  and unit $\hat{e}:\PSet(M)$ given by
  \[
    \hat{m}(p,q)(x)
      \defeq \exis{y:M}{\exis{z:M}{\ p(y) \land q(z) \land m(y,z)(x)}}
    \enspace,\quad
    \hat{e}(x) \defeq e(\ttt)(x) \enspace .
  \]
\end{theorem}

\begin{example}
  Since the inclusion $(-)_\ast:\Set\to\Rel$ maps products to tensors, every
  (commutative) monoid $(M;\mult,e)$ in $\Set$ yields a (commutative) monoid
  in $\Rel$ that, by promonoidal convolution
  (\cref{thm:promonoidal-convolution}), produces a
  (commutative) monoid $(\PSet(M),\hat\mult,\hat e)$ in $\Set$.
  In concrete terms,
  \[
    (p \mathrel{\hat{\mult}} q) (x)
    \jdgeq \exis{y:M}{\exis{z:M}{\ p(y) \land q(z) \land ( y\mult z \id x )}}
    \enspace,\quad
    \hat{e}\,(x) \jdgeq (e \id x)
    \enspace .
  \]
\end{example}

\noindent
The universal property of the free commutative monoid lifts from $\Set$ to
$\Rel$.
\begin{theoremrep}
  \label{RelFCM}
  For every set $A$, the set $\M(A)$ with multiplication
  $m\defeq(\mappend_A)_\ast:\M(A)\otimes\M(A)\proarrow\M(A)$ and unit
  $e\defeq(\lam{x:\unit}\mnil)_\ast :\unit\proarrow\M(A)$ equipped with
  $(\eta_A)_\ast : A\proarrow \M[A]$ is the free commutative monoid on $A$ in
  $\Rel$.
\end{theoremrep}

\begin{proof}
  For a commutative monoid $M$ in $\Rel$ and $f : A \proarrow M$, one needs to
  define a unique homomorphic extension $\M[A] \proarrow M$. Using that
  $\PSet(M)$ is a commutative monoid in $\Set$, this is given by $\extend{f}:
    \M[A]\to\PSet(M)$. The rest of the argument is established by calculation.
\end{proof}

\noindent
We turn our attention to the dual algebraic structure of commutative comonoid.

\begin{definition}[Commutative comonoid] 
  A \emph{commutative comonoid} in $\Rel$ is a set $K$ with: a
  \emph{comultiplication} $w : K \proarrow K \otimes K$ and a \emph{counit}
  $k : K \proarrow \unit$ satisfying the following commutative diagrams (up to
  homotopy):\\
  \begin{minipage}{\textwidth}
    \[
      \begin{tikzcd}[row sep = small, column sep = small]
        \unit \otimes K && K && K \otimes \unit
        \\ \\
        && K \otimes K &&
        \arrow["{w}", "\shortmid" marking, from=1-3, to=3-3]
        \arrow[swap, "{k \otimes K}", "\shortmid" marking, from=1-1, to=3-3]
        \arrow["{K \otimes k}", "\shortmid" marking, from=1-5, to=3-3]
        \arrow["{\eqv}", "\shortmid" marking, from=1-1, to=1-3]
        \arrow[swap, "{\eqv}", "\shortmid" marking, from=1-5, to=1-3]
      \end{tikzcd}
      \qquad
      \begin{tikzcd}[row sep = small, column sep = small]
        {K} && {K \otimes K} \\
        \\
        {K \otimes K} && {K \otimes K \otimes K}
        \arrow["{w}", "\shortmid" marking, from=1-1, to=1-3]
        \arrow["{w}"', "\shortmid" marking, from=1-1, to=3-1]
        \arrow["{K \otimes w}", "\shortmid" marking, from=1-3, to=3-3]
        \arrow["{w \otimes K}"', "\shortmid" marking, from=3-1, to=3-3]
      \end{tikzcd}
      \qquad
      \begin{tikzcd}[row sep = small, column sep = small]
        && {K \otimes K} \\
        {K} \\
        && {K \otimes K}
        \arrow["{w}", "\shortmid" marking, from=2-1, to=1-3]
        \arrow["{w}"', "\shortmid" marking, from=2-1, to=3-3]
        \arrow["\simeq\,"', "\,c", "\shortmid" marking, from=1-3, to=3-3]
      \end{tikzcd}
    \]
  \end{minipage}
  where $c\defeq(\langle\pi_2,\pi_1\rangle)_*$ is the symmetry isomorphism.
\end{definition}

\begin{example}
  Each set has a unique product commutative comonoid structure in $\Set$.
  This is preserved by the inclusion $(-)_\ast:\Set\to\Rel$ giving a
  commutative comonoid structure on the set in $\Rel$. This is the unique
  commutative comonoid structure on $\nulltype$ and $\unit$.
  In $\Rel$, by self-duality, every (commutative) monoid induces a (commutative)
  comonoid, and vice versa.
\end{example}

\begin{definition}[Homomorphism]
  For commutative comonoids $(K;w,k)$ and $(K';w',k')$, a relation
  $r : K \proarrow K'$ is a homomorphism whenever
  $w' \comp r \id (r \otimes r) \comp w$ and $k' \comp r \id k$.
\end{definition}

\begin{corollaryrep}
  For every set $A$,
  $\epsilon_A\defeq((\eta_A)_*)^\dagger:{\M[A]\proarrow A}$ is the cofree
  commutative comonoid on $A$ in $\Rel$; that is, for a commutative comonoid
  $K$, every relation $r : K \proarrow A$ has a unique homomorphic coextension
  $\coextend{r} : K\proarrow \M(A)$ over $\epsilon_A$.
\end{corollaryrep}

\begin{proof}
  This follows from~\cref{RelFCM} by self duality of $\Rel$. For a commutative
  comonoid $(K,w,k)$ in $\Rel$ and $r: K\proarrow A$, the homomorphic
  coextension $\coextend r : K \proarrow \M(A)$ is
  $(\extend{(r^\dagger)})^\dagger$ where the homomorphic extension is taken with
  respect to the promonoidal convolution on the commutative monoid
  $(K;w^\dagger,e^\dagger)$.
\end{proof}

\begin{theorem} \label{thm:RelLLmodel}
  The compact closed category $\Rel$ is a model of linear logic. The linear
  exponential comonad structure is as follows:
  \begin{description}
    \item comonad structure\\[-8mm]
          \begin{align*}
            \delta_A   & \defeq ((\mu_A)_*)^\dagger
            : \M(A) \proarrow \M(\M(A)) 
            \\
            \epsilon_A & \defeq ((\eta_A)_*)^\dagger
            : \M(A) \proarrow A 
          \end{align*}

    \item monoidal structure\\[-8mm]
          \begin{align*}
            \varphi_{A,B} & \defeq \coextend{(\epsilon_A\otimes\epsilon_B)}
            : \M(A) \otimes \M(B) \proarrow \M(A \otimes B)
            \\
            \phi          & \defeq \coextend{(\mathrm{id}_\unit)}
            : \unit \proarrow \M(\unit)
          \end{align*}
    \item commutative comonoid structure\\[-8mm]
          \begin{align*}
            w_A & \defeq ((\append_A)_*)^\dagger
            : \M(A) \proarrow \M(A)\otimes \M(A)
            \\
            k_A & \defeq ((\lam{x:\unit}\nil)_*)^\dagger: \M(A) \proarrow\unit
          \end{align*}
  \end{description}
\end{theorem}

\begin{proposition}
We have the following characterisation of the linear exponential monoidal
structure:
\[
\varphi = \big(\langle \M(\pi_1) , \M(\pi_2) \rangle_*\big)^\dagger
\enspace,\quad
\phi = \big(\langle\,\rangle_*\big)^\dagger
\enspace.
\]
\end{proposition}

The finite-multiset construction is canonically a strong symmetric monoidal
endofunctor on $\Rel$ from the biproduct monoidal structure to the tensor
monoidal structure (recall~\cref{prop:SeelyIso,prop:RelDaggerCompact}); the
monoidal isomorphisms are the Seely (or storage)
isomorphisms~\cite{seelyLinearLogicAutonomous1989}:
\begin{equation} \label{RelationalSeelyIsos}
  \M(A) \otimes \M(B) \stackrel\simeq\proarrow \M(A+B)
  \enspace,\quad
  \unit \stackrel\simeq\proarrow \M(\nulltype)
  \enspace.
\end{equation}


%% file: differentiallinearlogic.tex
\section{Relational model of differential linear logic}
\label{sec:DifferentialStructure}

We now give a formalisation of the relational model of Ehrhard and Regnier's
differential linear logic (see, for
instance,~\cite{ehrhardIntroductionDifferentialLinear2018a}).
We need to first establish combinatorial properties of subsingleton multisets.
As we will see in~\cref{sec:MultisetEqualitySection}, these structural
properties are also needed to characterise the equality type of finite
multisets.

\input{subsingletons}

\subsection{Differential structure}
\label{subsec:DifferentialStructure}

Differential structure in categorical models of linear logic can be described
in a variety of forms, see for
instance~\cite{bluteDifferentialCategoriesRevisited2020,fioreDifferentialStructureModels2007a,ehrhardIntroductionDifferentialLinear2018a}.
Here, we follow the axiomatisation of
Fiore~\cite{fioreDifferentialStructureModels2007a} in the context of models
with biadditive structure (that is, biproducts) as in
\cref{sec:relational-classical}.

In $\Rel$, this differential structure consists of a natural transformation
\emph{creation
  map}~\cite[Definition~4.3]{fioreDifferentialStructureModels2007a}:
\[
  \eta_A: A \proarrow {\M(A)}
\]
subject to three laws as follows:
\[\begin{tikzcd}
    & \M(A)
    \arrow[dr,"\shortmid"description]\arrow[dr,"\epsilon"]
    &
    \\
    A
    \arrow[ru,"\shortmid"description]\arrow[ru,"\eta"]
    \arrow[rr,"\shortmid"description]\arrow[rr,"\mathrm{id}"swap]
    & & A
  \end{tikzcd}
  %
  \qquad\qquad
  %
  \begin{tikzcd}
    A
    \arrow[r,"\shortmid"description]\arrow[r,"\eta"]
    \arrow[d,"-"description]\arrow[d,"\eqv"']
    & \M(A)
    \arrow[r,"\shortmid"description]\arrow[r,"\delta"]
    &
    \M^2(A)
    \\
    A\otimes \unit
    \arrow[r,"\shortmid"description]\arrow[r,"\eta\otimes e" swap]
    & \M(A)\otimes\M(A)
    \arrow[r,"\shortmid"description]\arrow[r,"\eta\otimes\delta" swap]
    & \M^2(A)\otimes\M^2(A)
    \arrow[u,"-"description]\arrow[u,"m" swap]
  \end{tikzcd}
  \]
  \[
  \begin{tikzcd}
    A\otimes\M(B)
    \arrow[r,"\shortmid"description]\arrow[r,"\eta\otimes\mathrm{id}"]
    \arrow[d,"-"description]\arrow[d,"\mathrm{id}\otimes\epsilon"swap]
    & \M(A)\otimes\M(B)
    \arrow[d,"-"description]\arrow[d,"\varphi"]
    \\
    A\otimes B
    \arrow[r,"\shortmid"description]\arrow[r,"\eta" swap]
    & \M(A\otimes B)
  \end{tikzcd}\]
This, as established
in~\cite[Section~4~\emph{\S\,Differentiation}]{fioreDifferentialStructureModels2007a}
(see also~\cite{bluteDifferentialCategoriesRevisited2020}) induces a
\emph{differential category}~\cite{bluteDifferentialCategories2006} structure.

In elementary terms, the above diagrams amount to the properties below, that
follow from the results of~\cref{subsec:singletons}.
\begin{proposition}
  \leavevmode
  \begin{enumerate}
    \item For $a,a':A$, we have
          \( a \id a' \Longleftrightarrow [a] \id [a']
          \).

    \item For $a: A$ and $s: \M^2(A)$, we have
          \(
            [a] \id \mu(s)
            \ \Longleftrightarrow \
            \exis{ t:{\M(\M(A))} } \ \mu(t)\id\nil \wedge \sing{a} \cons t \id s
          \).

    \item
          For $a:A$, $bs:\M(B)$, and $ps:\M(A\times B)$,\\[-2mm]
          \[
            [a]\id\M(\pi_1)(ps) \wedge bs\id\M(\pi_2)(ps)
            \ \Longleftrightarrow  \
            \exis{ b:B } \ bs \id [b] \wedge \big[(a,b)\big] \id ps
            \enspace .\]
  \end{enumerate}
\end{proposition}
\noindent
To summarise:
\begin{theorem}
  The category $\Rel$ is a model of differential linear logic.
\end{theorem}


%% file: subsingletons.tex
\subsection{Subsingleton multisets}
\label{subsec:singletons}


We start by describing empty and singleton multisets. These can be
characterised by their lengths (recall~\cref{ex:length}); namely, a multiset
$as$ is empty if it has length 0, $\length[as] \id 0$, and it is a singleton
if it has length 1, $\length[as] \id 1$. Alternatively, one can say that a
multiset is empty if it is equal to $\nil$ and it is a singleton if it is
equal to a one-element multiset. We will show that these are equivalent
notions.

\begin{definition}
  For $as:\M[A]$, we define~$\isEmpty[as] \defeq (as \id \nil)$
  and~$\isSing[as] \defeq \dsum*{a:A}{(as \id \sing{a})}$.
\end{definition}

\noindent
Since the equality type of $\M[A]$ is a proposition, it follows that
$\isEmpty[as]$ is a proposition. It is easy to see that this is equivalent to
being of length 0.

\begin{proposition}
  For $as : \M[A]$, $\isEmpty[as] \Leftrightarrow (\length[as] \id 0)$.
\end{proposition}

\noindent
It follows that free commutative monoids are
\emph{conical}~\cite{wehrungTensorProductsStructures1996}, that is, $\nil$ is
the only invertible element:
\begin{corollary}[Conical-monoid relation]
  For $as, bs : \M[A]$, $as\append bs\id\nil \Leftrightarrow as\id bs\id\nil$.
\end{corollary}

\noindent
Note that the dependent product defining $\isSing[as]$ is not truncated and
hence this type is not obviously a proposition. We can nevertheless show that
there is a \emph{unique choice} for such a witness, making it a proposition.
For length-one multisets, one can construct a unique choice function that
extracts the only element by induction on the structure of $\MSet[A]$.  This
allows us to establish the following results.  For details, we refer the
reader to the supplementary material in~\cref{prop:sing-uc-apx}.

\begin{propositionrep}
  \label{prop:sing-uc}
  Let $A$ be a set.
  \begin{enumerate}
    \item There is a unique choice function
          $\term{uc} : (as : \M[A]) \to (\length[as] \id 1) \to \isSing[as]$.
    \item The universal map $\eta: A \to \M[A]$ is an embedding; that is,
          $\term{ap}_{\eta} : x \id y \to \sing{x} \id \sing{y}$ is an
          equivalence.
    \item For $as:\M(A)$, $\isSing[as]$ is a proposition.
    \item For $as : \M[A]$, we have that $\term{uc}(as)$ is an equivalence.
  \end{enumerate}
\end{propositionrep}

\begin{proof}
  \begin{enumerate}
    \item The case for $\nil$ is eliminated since it has length 0. Knowing that
          $a \cons as$ has length 1 implies that $as$ has length 0 and is hence
          empty. Finally, $\swap$ only acts on multisets of length at least 2,
          hence the case for $\swap$ gets eliminated.
    \item Since $A$ and $\M[A]$ are sets, we only need to show that $\eta$ is
          injective, that is, $\sing{a} \id \sing{b}$ implies $a \id b$. Since
          these are length-one multisets, we simply apply the $\term{uc}$
          function.
    \item Follows from $\eta$ being injective.
    \item Since they are both propositions, we simply need an inverse map to
          $\term{uc}$, which we get by applying $\length$.
  \end{enumerate}
\end{proof}

\noindent
Using the above, we show that the type of singleton (or, equivalently,
length-one) multisets is equivalent to the underlying set.

\begin{propositionrep}
  For a set $A$,
  \[\textstyle
    A
    \ \eqv \
    \dsum*{as:\M[A]}{(\length[as] \id 1)}
    \ \eqv \ \dsum*{as:\M[A]}{\isSing[as]}
    \enspace.
  \]
\end{propositionrep}

\begin{proof}
  Follows from~\cref{prop:sing-uc} using the unique choice function.
\end{proof}

\noindent
Finally, we establish structural properties of singleton multisets arising
from concatenations and projections. These \emph{combinatorial properties}
will play a central role in~\cref{subsec:DifferentialStructure}.

\begin{propositionrep}
  Let $A$ and $B$ be sets.
  \begin{enumerate}
    \item
          For $s: \M(\M(A))$ and $a:A$, we have
          \(
          \mu(s) \id [a]
            \Leftrightarrow
            \exis{t:{\M(\M(A))}}{ \mu(t)\id\nil \wedge \sing{a} \cons t \id s}
          \).

    \item
          For $t:\M(A\times B)$ and $a:A$, we have
          \(
          \M(\pi_1)(t) = [a] \,\Leftrightarrow\, \exis{b:B}{t = \sing{(a,b)}}
          \).

    \item
          For $as, bs:\M(A)$ and $a:A$, we have
          \(
          as\append bs \id \sing{a}
            \Leftrightarrow
            (as \id \sing{a} \land bs \id \nil) \lor (as \id \nil \land bs \id
            \sing{a})
          \).
  \end{enumerate}
\end{propositionrep}

\begin{proof}
  Since these are equivalences of propositions, one simply needs to write down
  both the maps. Going from right to left is easy and just follows by
  calculation. To go from left to right, we use the characterisation of the path
  space of singleton multisets and the unique choice function
  from~\cref{prop:sing-uc}.
\end{proof}


%% file: equality.tex
\section{Path space of finite multisets}
\label{sec:MultisetEqualitySection}

An application of our development is an understanding of multiset equality.
We show that free commutative monoids are refinement
monoids~(\cref{prop:RefinementMonoid}) and, from there, characterise the path
space of free commutative monoids for non-empty multisets by means of a
\emph{commutation relation}~(\cref{thm:CommutationRelation}).  Then, we
further use this to characterise the path space of
finite multisets~(\cref{thm:PathSpace}) and to give a deduction system for
multiset equality~(\cref{subsec:DeductionSystem}).

\subsection{Commutation relation}
\label{subsec:CommutationRelation}

Two non-empty multisets of the form $x \cons xs$ are equal if they are either
equal point-wise (by the congruence of $\term{cons}$) or are equal up to a
permutation. Instead of explicitly working with permutations, we will
characterise this equality using a \emph{commutation relation} that stems from
the theory of creation/annihilation operators associated with the
finite-multiset construction seen as a combinatorial \emph{Fock space} as
developed by
Fiore~\cite{fioreDifferentialStructureModels2007a,fioreAxiomaticsCombinatorialModel2015}.

The following diagrams, involving the Seely
isomorphisms~(\ref{RelationalSeelyIsos}), commute (up to homotopy) 
\[\begin{tikzcd}[column sep = 1em, row sep = 1em]
    {\M(A)} \otimes {\M(A)}
    \arrow[dr,"\shortmid"description]\arrow[rd,"m"swap]
    \arrow[rr,"\shortmid"description]\arrow[rr,"\eqv"]
    &&
    \M(A+A)
    \arrow[dl,"\shortmid"description]\arrow[dl,"\M(\nabla)"]
    \\
    & \M(A)\arrow[dl,"\shortmid"description]\arrow[dl,"w"swap]
    \arrow[dr,"\shortmid"description]\arrow[dr,"\M(\Delta)"] &
    \\
    {\M(A)} \otimes {\M(A)} & &
    \arrow[ll,"\shortmid"description]\arrow[ll,"\eqv"]
    \M(A+A)
  \end{tikzcd}
  \qquad\qquad\qquad
  \begin{tikzcd}[column sep = 1em, row sep = 1em]
    \unit
    \arrow[dr,"\shortmid"description]\arrow[rd,"e"swap]
    \arrow[rr,"\shortmid"description]\arrow[rr,"\eqv"]
    &&
    \M(\nulltype)
    \arrow[dl,"\shortmid"description]\arrow[dl,"\M(u)"]
    \\
    & \M(A)\arrow[dl,"\shortmid"description]\arrow[dl,"k"swap]
    \arrow[dr,"\shortmid"description]\arrow[dr,"\M(n)"] &
    \\
    \unit & & \arrow[ll,"\shortmid"description]\arrow[ll,"\eqv"]
    \M(\nulltype)
  \end{tikzcd}\]
and one can use the symmetric monoidal coherence laws to transport the
biproduct commutative bialgebra structure through $\M$ as follows.  We refer
the reader
to~\cite{fioreDifferentialStructureModels2007a,fioreAxiomaticsCombinatorialModel2015}
for this theory.

\begin{proposition} \label{prop:BialgebraStructure}
  The biproduct commutative bialgebra structure $(\nabla,\Delta,u,n)$ on a set
  $A$ in $\Rel$ transfers to one on $\M(A)$ as $(m,w,e,k)$ for $(m,e)$ and
  $(w,k)$ respectively defined as in \cref{RelFCM,thm:RelLLmodel}.
\end{proposition}
%
%
\noindent
In particular, the bialgebra identity~(\ref{BiproductBialgebraLaw}) transports
to the one below:
\[\begin{tikzcd}
    \M(A)\otimes\M(A) & \M(A) & \M(A)\otimes\M(A)
    \\
    \M(A)\otimes\M(A)\otimes\M(A)\otimes\M(A) &&
    \M(A)\otimes\M(A)\otimes\M(A)\otimes\M(A)
    \arrow["{m}", "\shortmid"marking, from=1-1, to=1-2]
    \arrow["{w}", "\shortmid"marking, from=1-2, to=1-3]
    \arrow["{w\otimes w}"swap, "\shortmid"marking, from=1-1, to=2-1]
    \arrow["{m\otimes m}"swap, "\shortmid"marking, from=2-3, to=1-3]
    \arrow["{\idfun[\M(A)]\otimes c\otimes\idfun[\M(A)]}"swap,
    "\shortmid" marking, from=2-1, to=2-3]
  \end{tikzcd}\]
where $c$ is the symmetry isomorphism.
Spelling this out in elementary terms, we have that free commutative monoids
satisfy the \emph{Riesz refinement
  property}~\cite{dobbertinRefinementMonoidsVaught1983}.
\begin{proposition}[Refinement-monoid relation]
  \label{prop:RefinementMonoid}
  For $as,bs,cs,ds:\M(A)$,
  \[
    \begin{array}{c}
      as\append bs \id cs\append ds \\
      \Longleftrightarrow           \\
      \exis{ xs_1, xs_2, ys_1, ys_2 :\M(A) }\
      (as \id xs_1 \append xs_2)
      \ \land \
      (bs \id ys_1 \append ys_2)
      \ \land \
      (xs_1 \append ys_1 \id cs)
      \wedge
      (xs_2 \append ys_2 \id ds) \ .
    \end{array}
  \]
\end{proposition}
\noindent
Using this identity, we obtain the characterisation below of the equality type
for non-empty finite multisets. 
\begin{theoremrep}[Commutation relation]
  \label{thm:CommutationRelation}
  For a set $A$, $a,b:A$, and $as,bs:\M[A]$,
  \[
    a \cons as \id b \cons bs
    \enspace \Leftrightarrow \enspace
    (a \id b \land as \id bs)
    \, \lor \,
    (\exis{ cs:\M(A) }\ as \id b \cons cs \land a \cons cs \id bs)
    \enspace . \]
\end{theoremrep}

\begin{proof}
  This follows from the bialgebra law (refinement-monoid relation) and using
  the following sequence of manipulations which use the characterisation of
  singletons.
  \begin{enumerate}
    \item For $a, b : A$, and $as : \M[A]$, we have
          \[ a \cons as \id \sing{b} \Longleftrightarrow (a \id b) \land (as \id \nil)
            \enspace. \]
    \item For $a, b : A$, and $as, cs : \M[A]$, we have
          \[ \exis{xs : \M[A]} (a \cons xs \id \sing{b}) \land (as \id xs \append cs)
            \Longleftrightarrow (a \id b) \land (as \id cs)
            \enspace. \]
    \item For $a : A$, and $as, bs, cs : \M[A]$, we have
          \[ \begin{array}{c}
              (a \cons as) \id (bs \append cs) \\
              \Longleftrightarrow              \\
              \exis{xs : \M[A]} (a \cons xs \id bs) \land (as \id xs \append cs)
              \lor
              \exis{ys : \M[A]} (as \id bs \append ys) \land (a \cons ys \id cs)
              \enspace.
            \end{array} \]
  \end{enumerate}
\end{proof}

\noindent
The commutation relation can be thought of as consisting of a congruence rule
and a \emph{generalised}~$\term{swap}$ rule.  The latter can be written as a
deduction rule for multiset equality as follows:
\begin{equation}\label{CommRule}
  \begin{array}{c}
    as \id b \cons cs \qquad a \cons cs \id bs
    \\ \hline
    a \cons as \id b \cons bs
  \end{array}\ \comm
\end{equation}
As exemplified below, the $\comm$ rule allows one to generate and compose
$\swap$ operations.

\begin{example}
  \label{ex:comm}
  For $a, b, c : A$ and $cs, ds : \M(A)$,\\
  \begin{enumerate}
    \item \mbox{}\\[-5mm]
          \[\begin{prooftree}
              \infer0[$\refl$]{ b \cons cs \id b \cons cs }
              \infer0[$\refl$]{ a \cons cs \id a \cons cs }
              \infer2[$\comm$]{ a \cons b \cons cs \id b \cons a \cons cs }
            \end{prooftree}\]

    \item \mbox{}\\
          \[\begin{prooftree}
              \infer0[$\refl$]{ b \cons ds \id b \cons ds }
              \infer0[$\refl$]{ c \cons ds \id c \cons ds }
              \infer2[$\comm$]{ b \cons c \cons ds \id c \cons b \cons ds }
              \infer0[$\refl$]{ a \cons ds \id a \cons ds }
              \infer0[$\refl$]{ b \cons ds \id b \cons ds }
              \infer2[$\comm$]{ a \cons b \cons ds \id b \cons a \cons ds }
              \infer2[$\comm$]
              { a \cons b \cons c \cons ds \id c \cons b \cons a \cons ds }
            \end{prooftree}\]
  \end{enumerate}
\end{example}

To show that the commutation relation generates the path space of
finite multisets, we define an inductive relation on $\M[A]$ and establish an
equivalence with the equality type.
\begin{definition}
  For a type $A$, let $\ \rel_A\ : \M(A)\to \M(A)\to\UU$ be the inductive
  relation with the following constructors:\\[-8mm]
  \begin{align*}
    \nilcong  & : \nil \rel_A \nil                                                                                                              \\
    \conscong & : \{ a \ b : A \}\ \{ as \ bs : \M(A) \} \to (a = b) \to (as \rel_A bs) \to (a::as \rel_A b::bs)                                \\
    \commrule & : \{a\ b : A\}\ \{as\ bs\ cs : \M[A]\} \to (as \rel_A b \cons cs) \to (a \cons cs \rel_A bs) \to (a \cons as \rel_A b \cons bs)
  \end{align*}
\end{definition}
\noindent
Note that this relation is not valued in propositions, since there may be
multiple derivations of the same permutation.
\begin{example}
  \label{ex:commrel-dup}
  For $a:A$, the type $(\ a :: a :: \nil \rel_A a :: a :: \nil \ )$
  is inhabited by both of these terms:
  $\conscong(\refl,\conscong(\refl,\nilcong))$ and
  $\commrule(\,\conscong(\refl,\nilcong)\,,\,\conscong(\refl,\nilcong)\,)$.
\end{example}
\noindent
Hence, we propositionally truncate the relation. Then, a usual encode-decode
argument shows that this is equivalent to the path space of $\M[A]$.
\begin{proposition}
  For a set $A$ and $as, bs:\M(A)$, we have a soundness map $( as \rel_A bs )
    \to ( as \id_{\M(A)} bs )$ and a reflexivity map $(\, as: \M(A) \,) \to
    \Trunc[-1]{as \rel_A as}$.
\end{proposition}
\begin{theorem} \label{thm:PathSpace}
  For a set $A$ and $as, bs:\M(A)$, we have the characterisation
  $(\, as \id_{\M(A)} bs \,) \Leftrightarrow \Trunc[-1]{as \rel_A bs}$.
\end{theorem}


%% file: deduction-system.tex
\subsection{Multiset-equality deduction system}
\label{subsec:DeductionSystem}

The commutation relation of the previous section can be simply defined as a
relation on lists and, in the presence of congruence rules, this yields that
two lists are related iff they are equal when regarded as multisets. This
provides a new construction of $\M[A]$ as a set-quotient of $\List(A)$, which
we present next.

\begin{definition} \label{def:cList}
  For a type $A$, let $\ \rel_A\ : \List[A] \to \List[A] \to \UU$ be the
  inductive relation generated by the following constructors:
  \begin{align*}
    \nilcong  & : \nil \rel_A \nil                                                                                                              \\
    \conscong & : \{ a \ b : A \}\ \{ as \ bs : \List[A] \} \to (a \id b) \to (as \rel_A bs) \to (a::as \rel_A b::bs)                              \\
    \commrule & : \{a\ b : A\}\ \{as\ bs\ cs : \List[A]\} \to (as \rel_A b \cons cs) \to (a \cons cs \rel_A bs) \to (a \cons as \rel_A b \cons bs)
  \end{align*}
\end{definition}

\noindent
It is straightforward to show that the relation $\rel_A$ is reflexive and
symmetric, and a congruence with respect to $\append$.  It is however
non-trivial to show that it is transitive.

Note that the relation $\rel_A$ encodes permutations of the elements of the
related lists; alternatively, permutations on the finite set of indices of the
lists.  We have an equivalence
\[\textstyle
\vectorise
  \, : \, \List[A] 
  \ \eqv \
  \big(\dsum*{\ell:\Nat}{\Fin[\ell] \to A}\big)
  \, : \, \listify
\]
and we now define a new relation $\perm_A$ on the above representation of
lists by relating two listing functions if, and only if, there exists a
permutation of their indices that shuffles one into the other.

\begin{definition} \label{def:perm-rel}
For a type $A$, define:
  \begin{align*}
     & \textstyle
     {\blank}\perm_A{\blank} 
     : \big(\dsum*{\ell:\Nat}{\Fin[\ell] \to A}\big)
       \to \big(\dsum*{\ell:\Nat}{\Fin[\ell] \to A}\big)
       \to \UU 
     \enspace, \\[2mm]
     & (m , f) \perm_A (n , g) 
       \defeq 
       \ps{\phi : \Fin[m] \xrightarrow{\sim} \Fin[n]} 
       \times 
       \ps{f \id g \comp \phi}
     \enspace.
  \end{align*}
\end{definition}

\noindent
Since permutations are simply invertible functions, it is straightforward to
establish that this is an equivalence relation. We now show that the relation
$\perm_A$ \emph{extensionally} agrees with our relation~$\rel_A$. To this end,
we perform a translation between the two relations in the style of an NbE
(Normalisation by Evaluation) algorithm.

First, given a deduction tree for $\rel_A$, we compute the permutation it
encodes by means of an $\term{eval}$ function.  Second, given an inhabitant of
$\perm_A$ we reify it to a deduction tree by means of a $\term{quote}$
function (see~\cref{def:eval-quote-apx} for details).

\begin{definitionrep}\label{def:eval-quote}
For $as, bs: \List(A)$, we have 
\[
  \term{eval} : \ as \rel_A bs \to \vectorise(as) \perm_A \vectorise(bs)
\]
and, for $(m,f), (n,g): \big(\dsum*{\ell:\Nat}{\Fin[\ell] \to A}\big)$, we
have
\[
  \term{quote} : \ (m,f) \perm_A (n,g) \to \listify(m,f) \rel_A \listify(n,g)
  \enspace.
\]
\end{definitionrep}

\begin{proof}
  We define $\term{eval}$ by induction on $as \rel_A bs$. The cases for
  $\nilcong$ and $\conscong$ are trivial. For the $\commrule$ case, we need to
  calculate ${\vectorise(a \cons as)} \perm_A {\vectorise(b \cons bs)}$. We
  recursively calculate $\vectorise(as) \perm_A {\vectorise(b \cons cs)}$ and 
  ${\vectorise(a \cons cs)} \perm_A {\vectorise(bs)}$.  Then, by congruence,
  we have ${\vectorise(a \cons as)} \perm_A {\vectorise(a \cons b \cons cs)}$
  and ${\vectorise(b \cons a \cons cs)} \perm_A {\vectorise(b \cons bs)}$. We
  construct the permutation that swaps the first two elements to get 
  ${\vectorise(a \cons b \cons cs)} \perm_A {\vectorise(b \cons a \cons cs)}$.
  Finally, we compose the three to get 
  ${\vectorise(a \cons as)} \perm_A {\vectorise(b \cons bs)}$.

  To construct $\term{quote}$, we start with $(m,f) \perm_A (n,g)$; that is,
  we have a $\phi : \Fin[m] \eqv \Fin[n]$ such that $f = g \comp \phi$.
  Knowing $\phi$, we deduce that $m \id n$ and proceed by induction on $m:\Nat$.

  \begin{description}
    \item Case $
      \term{0}$: 
      Follows by $\nilcong$.
    \item Case $
      \suc\!(\term{0})$: 
      Since $\Fin[\suc\!(\term{0})] \eqv \Fin[\suc\!(\term{0})]$ is
      contractible, $\phi$ is the identity. Using this, we get a path between
      the head elements $f(\term{0}) \id g(\term{0})$ and then apply
      $\conscong$ using $\nilcong$.
    \item Case $
      \suc\!(\suc\!(\ell))$: 
      We use the decidable equality on $\Fin[\suc\!(\suc\!(\ell))]$ to test
      whether or not $\phi(\term0) \id \term0$.
          \begin{description}
            \item Case $\phi(\term0) \id \term0$: 
              We get a path between the elements 
              $f(\term0) \id g(\phi(\term0)) \id g(\term0)$.
              We also have 
              $(\suc\!(\ell),f\circ\suc) \perm_A (\suc\!(\ell),g\circ\suc)$
              and recursively obtain
              $\listify(\suc\!(\ell),f\circ\suc) 
               \rel_A 
               \listify(\suc\!(\ell),g\circ\suc)$.
              Finally, we combine them using $\conscong$.

            \item Case $\phi(\term0) \id \suc\!(k)$: 
              We need the following auxiliary definition: For
              $t:\Fin[\suc\!(i)]$, we let 
              $\widehat t: \Fin[i]\to\Fin[\suc\!(i)]$ be given by 
              $\widehat t(j) \id j$ for $j<t$ and by 
              $\widehat t(j)=\suc\!(j)$ for $j\geq t$.  
              We then have that 
              \[
                (\,\suc(\ell)\,,\,f\circ\suc\,)
                \perm_A
                (\,\suc(\ell)\,,\,g\circ\widehat{\phi(\term0)}\,)
              \]
              and therefore obtain
              \begin{equation}\label{E1}
                \begin{array}{rcll}
                  & & 
                \listify(\,\suc(\ell)\,,\,f\circ\suc\,)
                \\
                & \rel_A &  
                \listify(\,\suc(\ell)\,,\,g\circ\widehat{\phi(\term0)}\,)
                & \text{, by recursion}
                \\
                & \id &
                g(\term0)
                \cons
                \listify(\,\ell\,,\,g\circ\widehat{\suc\!(k)}\circ\suc\,)
                \\
                & \id & 
                g(\term0)
                \cons
                \listify(\,\ell\,,\,g\circ\suc\circ\,\widehat{k}\,)
              \end{array}
            \end{equation}
              On the other hand,
              \[
                \big(\, 
                  \suc\!(\ell) 
                  \, , \,
                  [\,\term0\mapsto g(\phi(\term0))
                     \mid
                     \suc\!(i)\mapsto g(\suc\!(\widehat k(i)))\,]
                \, \big)
                \ \perm_A \ 
                \big(\,\suc(\ell)\,,\,g\circ\suc \,\big)
              \]
              and we further obtain
              \begin{equation}\label{E2}
              \begin{array}{lll}
                & 
                f(\term0)\cons
                \listify(\, \ell \, , \, g\circ\suc\circ\,\widehat k \, )
                \\
                \id & 
                g(\phi(\term0))\cons
                \listify(\, \ell \, , \, g\circ\suc\circ\widehat k \, )
                \\
                \id & 
                \listify(\, 
                  \suc\!(\ell) 
                  \, , \,
                  [\, \term0\mapsto g(\phi(\term0))
                      \mid
                      \suc\!(i)\mapsto g(\suc\!(\widehat k(i))) \,]
                 \, )
                \\
                \rel_A & 
                \listify(\,\suc(\ell)\,,\,g\circ\suc \,)
                & \text{, by recursion}
               \end{array}
               \end{equation}
             Finally, an application of the $\commrule$ to $(\cref{E1})$ and
             $(\cref{E2})$ yields
             \[
                \listify(\,\suc\!(\suc(\ell))\,,\,f\,)
                \, \id \,
                f(\term0)\cons\listify(\,\suc\!(\ell)\,,\,f\circ\suc\,)
                \, \rel_A \,
                g(\term0)\cons\listify(\,\suc\!(\ell)\,,\,g\circ\suc \,)
                \, \id \,
                \listify(\,\suc\!(\suc\!(\ell))\,,\,g \,)
              \]
              as required.
          \end{description}
  \end{description}
\end{proof}

\noindent
Finally, we establish the transitivity of $\rel_A$ by translating two
composable trees in $\rel_A$ to a composable pair in $\perm_A$ and, after
using the transitivity of $\perm_A$, quoting back the result to a tree. 

\begin{lemmarep}
For a type $A$, $\rel_A$ is an equivalence relation on $\List[A]$ and a
congruence with respect to~$\append$.
\end{lemmarep}

\begin{proof}
  Transitivity follows by~\cref{def:eval-quote} and using the transitivity of
  the relation~$\perm_A$. The rest is straightforward to establish.
\end{proof}

\noindent
The relation $\rel_A$ is not valued in propositions, since there may be
multiple inhabitants that encode the same permutation of the underlying
elements (see~\cref{ex:commrel-dup}). We propositionally truncate it, defining
$xs \rel*_A ys \defeq \Trunc[-1]{xs \rel_A ys}$, before using it in as a
quotient.  The resulting effective quotient of lists satisfies the categorical
universal property of free commutative monoids.  

\begin{theorem}
  For a set $A$, the composite $A \to \List[A] \to \List[A]_{/{\rel*_A}}$ is
  the free commutative monoid on $A$.
\end{theorem}

\begin{corollary} \label{cor:ListQuotient} For every set $A$, we have the
  equivalence $\List[A]_{/{\rel*_A}} \eqv_{\type{CMon}}\, \FCM[A]$.
\end{corollary}

\subsection{Commuted-list construction}
\label{subsec:CommutedListConstruction}

As a final application of the commutation
relation~(\cref{thm:CommutationRelation}), we give a construction of another
HIT, $\NSet$~(\cref{def:commList}), for finite multisets that uses the
$\commrule$~(\ref{CommRule}) as a path constructor. This is an example of a
conditional path constructor, which corresponds to a quasi-equational (or
Horn) theory.

\begin{definition}[$\NSet$] \label{def:commList}
  For a type $A$, the
  1-HIT $\NSet[A]$ is given by the following point and path constructors:
  \begin{align*}
    \nil                  & : \NSet[A]                                        \\
    {\blank}\cons{\blank} & : A \times \NSet[A] \to \NSet[A]                  \\
    \comm                 & : \{a\;b : A\} \{as\;bs\;cs : \NSet[A]\}          \\
                          & \to (as \id b \cons cs ) \to (a \cons cs  \id bs) \\
                          & \to a \cons as \id b \cons bs                     \\
    \truncc               & : \isSet{\NSet[A]}
  \end{align*}
\end{definition}

\begin{toappendix}
  \begin{definition}[Induction principle for $\NSet$]
    For every type family $P : \NSet[A] \to \UU$ with the following data:
    \begin{align*}
      \nil*                  & : P(\nil)                                                       \\
      {\blank}\cons*{\blank} & : (a : A) \{as : \NSet[A]\} \to P(as) \to P(a \cons as)         \\
      \comm*                 & : \{a\;b : A\} \{as\;bs\;cs : \NSet[A]\}                        \\
                             & \to \{pas : P(as)\} \{pbs : P(bs)\} \{pcs : P(cs)\}             \\
                             & \to (p : as \id b \cons cs ) \to (pas \id_{p}^{P} b \cons* pcs) \\
                             & \to (q : a \cons cs \id bs) \to (a \cons* pcs \id_{q}^{P} pbs)  \\
                             & \to a \cons* pas \id_{\comm[p,q]}^{P} b \cons* pbs              \\
      \truncc*               & : (as : \NSet[A]) \to \isSet{P(as)}
    \end{align*}
    there is a unique function $f : {(as : \NSet[A])} \to P(as)$ with the following computation rules.
    \begin{align*}
      f(\nil)             & \jdgeq nil*                           \\
      f(a \cons as)       & \jdgeq a \cons* f(as)                 \\
      \apd{f}{\comm[p,q]} & \id \comm*[p,\apd{f}{p},q,\apd{f}{q}]
    \end{align*}
  \end{definition}
\end{toappendix}
\noindent
Unsurprisingly, this HIT also satisfies the universal property of free
commutative monoids, making it equivalent to the other presentations.

\begin{proposition}
  For every set $A$, $\NSet[A]$ is a commutative monoid.
\end{proposition}

\begin{theorem}
  For every set $A$, $\eta:A\to\NSet[A]$ is the free commutative monoid on $A$.
\end{theorem}


%% file: discussion.tex
\section{Discussion}
\label{sec:Discussion}

We have described various constructions of free commutative monoids, or
finite multisets, using set-truncated HITs 
and a set quotient. 
Our results show that a significant part of the combinatorics of
finite multisets can be reduced to constructions using the categorical universal
property and a few structural properties. Our main application is a constructive
formulation of the relational model of differential linear logic, encompassing
the theory of creation/annihilation operators associated to the combinatorial
Fock-space construction. From this, we obtained a commutation relation that we
used to characterise the path space of finite multisets and further provided a
deduction system for multiset equality. A preliminary version of our work was
presented in~\cite{choudhuryFinitemultisetConstructionHoTT2019}.

\subsection{Finite multisets}

The simplest encoding of finite multisets is lists up to permutation of their
elements. In this context, the relation of \cref{def:perm-rel} using
automorphisms on finite cardinals as permutations has been considered by
Altenkirch et al.~\cite{altenkirchDefinableQuotientsType} and
Nuo~\cite{liQuotientTypesType2015}; while
Danielsson~\cite{danielssonBagEquivalenceProofRelevant2012} has considered a
similar relation for bag equivalence of lists.
Multisets in type theory have also been studied by
Gylterud~\cite{gylterudMultisetsTypeTheory2020} from the point of view of
constructive foundations, considering how to generalise set-theoretic axioms
from sets to multisets. They take the definition of a set as a W-type and use
a different equality relation to obtain multisets.
Using HITs in HoTT/UF, Angiuli et
al.~\cite{angiuliInternalizingRepresentationIndependence2021} considered two
representations of finite multisets in Cubical Agda: one is the swapped-list
construction and the other is association lists encoding elements with their
multiplicity; these are shown equivalent. They are interested in the
representation independence of data structures and assume decidable equality on
the carrier set.


\subsection{Commutation axioms}

The definition of various (higher) algebraic structures in HoTT/UF using HITs
is an active line of research; for example, the problem of defining (free
higher) groups has been considered
by~\cite{krausFreeHigherGroups2018,buchholtzHigherGroupsHomotopy2018,altenkirchIntegersHigherInductive2020}.
For algebraic structures with commutation axioms, there are various techniques
which are known folklore.

Basold et al.~\cite{henningbasoldHigherInductiveTypes2017} and Frumin et
al.~\cite{fruminFiniteSetsHomotopy2018} consider Kuratowski-finite sets in
HoTT defined as free join semilattices using the universal-algebraic
construction.  Our universal-algebraic construction of free commutative
monoids (\cref{def:fcm}) is similar to their HIT~$\mathcal{K}(A)$ but lacking
the idempotence axiom. The swapped-list construction~(\cref{def:mset}, or
listed finite multisets) is similar to their HIT~$\mathcal{L}(A)$ (or listed
finite sets) but lacking the duplication axiom. Frumin et
al.~\cite[Theorem~2.8]{fruminFiniteSetsHomotopy2018} show that these two HITs
are equivalent by constructing an explicit equivalence. In contrast, we prove
the categorical universal property of free commutative monoids for each HIT,
thereby getting their equivalence for free.


Formalisations of free abelian monoids and free abelian groups in HoTT/UF also
appear in the UniMath~\cite{UniMath},
HoTT~\cite{bauerHoTTLibraryFormalization2017}, and
HoTT-Agda~\cite{hott-in:agda} libraries; they use various forms of commutation
axioms.
In UniMath, free abelian monoids are defined by quotienting free monoids
(lists) by the relation
$g \relop h \defeq \exis{x,y}{\, g\id x \append y \land y \append x \id h}$.
In the HoTT (Coq) library, abelianisation is defined using a homotopy
coequaliser (or, equivalently, a HIT) where the commutation axiom is
essentially defined as
$\fora{x,y,z}{\, x \cdot (y \cdot z) \id x \cdot (z \cdot y)}$.
In HoTT-Agda, free abelian groups are defined by abelianising the free group,
essentially using the commutation relation
$l_{1} \append l_{2} \id l_{2} \append l_{1}$ on lists.
We use the obvious commutation axiom in the universal-algebraic construction,
and adjacent transpositions ($\swap$) in the swapped-list construction, which is
enough to prove the categorical universal property and the structural properties
that we use in our applications. The commutation relation and the conical
refinement monoid relations that we prove in~\cref{subsec:CommutationRelation}
have not been considered before.

\subsection{Higher Inductive Types}

All three HITs, $\FCM$ (\cref{def:fcm}), $\MSet$ (\cref{def:mset}), and
$\NSet$ (\cref{def:commList}) are recursive HITs, since the point and path
constructors refer to points in the type.  However, $\NSet$ is a recursive HIT
with a conditional path constructor $\comm$, since it refers to the path space
of the type. These definitions of HITs are accepted by (the current version
of) Cubical Agda~\cite{vezzosiCubicalAgdaDependently2019}. However, it is
unclear whether the known schemas for HITs in cubical type theories can encode
conditional path constructors. Fiore et
al.~\cite{fioreQuotientsInductiveTypes2022} consider $\MSet$ in their
quotient-inductive types framework and mention $\NSet$ as an example that goes
beyond it.

\subsection{Relational model of linear logic}

Standard presentations of the relational model of linear logic, see
e.g.~\cite[Section 4]{bucciarelliNotEnoughPoints2007},~\cite[Section
2.1]{ehrhardScottModelLinear2012}, ~\cite[Section
III.B]{lairdWeightedRelationalModels2013}, or~\cite[Section
II.A]{ongQuantitativeSemanticsLambda2017}, are typically written in informal
mathematical vernacular and often implicitly assume further structure on sets
when manipulating the finite-multiset construction. For instance, finite
multisets may be listed, thereby implicitly relying on an underlying order, and
the permutation invariance of constructions may be glossed over; or they may be
represented as finitely supported $\Nat$-valued functions sometimes relying on
the decidability of equality and without dwelling on notions of finiteness. On
the other hand, our constructive treatment of the subject has forced us to be
completely rigorous in all respects.

\subsection{Groupoidification}

This work is about 0-truncated finite multisets, which only allow elimination
into sets. This is somewhat restrictive. 
The preliminary presentation of our
work~\cite{choudhuryFinitemultisetConstructionHoTT2019} included a proposal for
extensions to free symmetric monoidal groupoids and the construction of the
bicategory of generalised species for groupoids. In particular, generalising the
above HITs to groupoids requires higher path constructors; for example, $\MSet$
requires a higher $\term{hexagon}$ path constructor to establish the coherence
of $\swap$.

It is folklore that homotopy finite-multi-types in HoTT should be
given by the construction
\( \M[A] \defeq \dsum*{X:\UU_{\Fin}}{\pi_{1}(X) \to A} \) where
\(
\UU_{\Fin} \defeq \dsum*{X:\UU}{\dsum*{n:\Nat}{\Trunc[-1]{X \id \Fin[n]}}}
\)
(see, for instance,~\cite{buchholtzGenuinePairsTrouble2021}).
It is generally believed that this should satisfy the universal property of
the free symmetric monoidal ($\infty$)~groupoid on the ($\infty$)~groupoid
$A$, but it is unknown how to prove this internally in HoTT.

Truncated to 1-groupoids, the construction of HITs that satisfy the universal
property of free symmetric monoidal groupoids has been considered by
Piceghello~\cite{piceghelloCoherenceSymmetricMonoidal2019,piceghelloCoherenceMonoidalSymmetric2021a}.
The equivalence of the free symmetric monoidal groupoid on one generator with
$\UU_{\Fin}$ in HoTT has been considered by Choudhury et
al.~\cite{choudhurySymmetriesReversibleProgramming2022a,choudhuryArtifactSymmetriesReversible2021}
in the context of building fully-abstract models for reversible programming
languages~\cite{choudhurySymmetriesReversibleProgramming2022a,caretteReversibleProgramsUnivalent2018}.
We believe that it is possible to apply these techniques to generalise the work
presented in this paper to groupoids.


\subsection{Formalisation}
\label{subsec:formalisation}

As mentioned in the introduction, we have a partial formalisation of the results
in this paper using Cubical Agda. The formalisation is around 3000 lines of code
and is available
at~\url{https://github.com/vikraman/generalised-species/tree/master/set}. The
status of the formalisation is summarised in~\cref{tbl:formalisation} in the
appendix. Some key highlights and differences with the paper are:


\begin{itemize}
  \item In the definition of commutative monoid structure, we drop one of the
        unit axioms since it can be derived from commutativity.
  \item The Seely isomorphism and commutative monad structure can be proved
        using the categorical universal property; however, we sometimes work with the
        type-theoretic induction principle of $\MSet$ (\cref{def:mset:ind}
        and~\cref{lem:mset:prop-ind}) because it is easier to formalise.
  \item When working with lists and permutations, we use different, more
        convenient representations of $\List$ and $\Fin$, these are equivalent
        and the results can be transported.
  \item The HIT for the swapped-list construction was merged into the Cubical
        Agda library~\cite{StandardLibraryCubical}, and various other results
        about finite multisets and free commutative monoids have been formalised
        by contributors to the library.
\end{itemize}

\begin{toappendix}
At the time of writing, the following results have been fully formalised, while
the remaining ones have only been partially formalised.

\begin{table}[ht]
  \centering
  \begin{tabular}{|l|l|}
    \hline
    {\tt set.CMon}
    & universal-algebraic construction (\cref{subsec:UniversalAlgebraConstruction}) \\
    \hline
    {\tt set.MSet}
    & swapped-list construction (\cref{subsec:FiniteMultisetConstruction}) \\
    \hline
    {\tt set.NSet}
    & commuted-list construction (\cref{subsec:CommutedListConstruction}) \\
    \hline
    {\tt set.QSet}
    & quotiented-list construction (\cref{subsec:CommutedListConstruction}) \\
    \hline
    {\tt SIP} & SIP for commutative monoids (\cref{prop:sip-cmon}) \\
    \hline
    {\tt MSet.Length}, {\tt MSet.Paths} & subsingleton multisets (\cref{subsec:singletons}) \\
    \hline
    {\tt MSet.Cover} & commutation relation and path space (\cref{subsec:CommutationRelation}) \\
    \hline
    {\tt Monoidal} & Seely isomorphism (\cref{prop:SeelyIso}) \\
    \hline
    {\tt Comm} & commutative monad structure (\cref{prop:CommutativeMonad}) \\
    \hline
    {\tt Power}, {\tt nPSh}, {\tt PSh} & power objects and presheaves (\cref{subsec:PowerObjects}) \\
    \hline
    {\tt hRel} & category of relations (\cref{subsec:CategoryRel}) \\
    \hline
    {\tt Diff} & differential structure (\cref{subsec:DifferentialStructure}) \\
    \hline
  \end{tabular}
  \caption{Status of formalised results}
  \label{tbl:formalisation}
\end{table}

\end{toappendix}
